# Individualistic networks and networking individuals in *Tango Argentino*

Céline Kuttler and Ralf Blossey

Interdisciplinary Research Institute, c/o IEMN, Cité Scientifique - Avenue Poincaré BP 69, F-59652 Villeneuve d'Ascq, France

**We have performed a simulation study of the social network arising from dancing partner selection in *Tango Argentino*. *Tango Argentino* is a famous 'intellectual' dance which combines individualistic behaviour with consensual social rules, generating complex patterns of network behaviour in time. We quantify these patterns by time-dependent degree distributions and clustering coefficients for the network structure, and by monitoring the evolution of the individual dancers' skill. In particular, we have investigated how successful new dancers are in entering the network under the promotion of established network members. Our approach allows us to predict the success of mentoring in a social network.**

Structural features of complex networks are currently a topic of intense research interest[1], triggered by recent insights on small-world behaviour[2] and scale-free network topologies arising from network growth[3]. A special class of complex networks arises from social interactions, with examples being science collaboration networks, movie actor networks, the WWW, but also certain animal networks[1,2,4,5]. All of them can be described by a graph structure of nodes (these describe the network members) and links (representing the social interactions of the members)[6]. Social networks are special as they show specific clustering effects, called assortative mixing[7], i.e. the existence of positive degree correlations between adjacent network nodes. Clustering is found to be very strong in these networks as compared to other complex networks[8]. The origin of the clustering appears to be the division of social networks into communities of acquaintances.

The division of social networks into communities of acquaintances already points to the fact that for a realistic description of social networks individual characteristics ultimately have to enter. Rather than developing a network model and applying it to data taken from specific data bases, which for the inclusion of individual features are not readily available, we had to opt for a different route. We looked for a 'real' social network which displays both community and individual features. Our example, the dancer network of Tango Argentino[9], which we describe here and model below, is a model social system in which the structural features of the network and individual dynamics are strongly coupled. It can be understood as a network of N = F + M dancers, with F female (called *tangueras*), and M male (called *tangueros*), participating at a dance event (called *milonga*). A dance between a male and a female dancer establishes a link between them; when the same dancers perform additional dances, their link is reweighted according to the number of these dances. Consequently, the node degree of each node i can be characterised by a time-dependent value $k_i(t)$, describing the number of established connections, and $D_i(t)$ for the number of dances performed, giving rise to degree distributions $P(k,t)$ and the distribution of the total number of dances, $P(D,t)$.

*Tango Argentino* networks constitute social groups which enforce a compromise of individual behaviour with social interaction rules based on a general consensus among its aficionados. A typical tango community consists of approximately not more than N = 100 dancers. Tango dancing is mostly organized by non-commercial associations or clubs (dance salons, also called milongas). Its networks are characterized by their group orientation and openness: despite the fact that tango is a social dance for couples, the communities' spirit is group oriented. Individuals are urged not to monopolize another on the dance floor, regardless of existing private or romantic relationships. An ideal dancing event is one in which each partner had several other dancing partners. Moreover, tango communities are eager to integrate new members. As a consequence of this strong social orientation of the community of tango dancers, a well-defined agreement on social rules of behaviour exist, in stark contrast to other dancing communities. The fact that *Tango Argentino* is not a predominantly individualistic dance is a prerequisite for any nontrivial dynamics to occur.

In modelling the network, we have drawn upon a recent empirical study providing data on the composition of tango communities in Germany[10]. The basic empirical findings relevant for our modelling purpose are: a) tango dancers are tightly bound to their community: about 40 % of the participants dance more than once a week, 30 % at least once, and about 15 % several times a month (rounded figures); b) tango communities have a high educational level (in Germany, about 75 % are university graduates, to be compared with a national average of about 14 %); c) the age distribution is wide (20-80 y) with an average of roughly 40 years; d) about 40 % of the dancers are married. Consequently, the focus of interest in participation is driven by social interaction via dancing, with romantic ambitions being less relevant. The tango network membership can thus be considered a rather homogeneous social group – a feature which simplifies the modelling effort considerably while still being faithful to the tango community.

The selection of a dancing partner means placing a link between tanguero and tanguera according to the following rules that can be formalized: a) the body height of the male dancer must be larger than that of the female dancer by a given amount, as the male dancer is preferred to be taller than the female (although also not by too much, so that also an upper limit exists); b) the dancing skills of male and female dancers must not differ by too much; c) dancers are not allowed to monopolize each other: only a certain number of common dances are allowed during the dancing event; d) the same partners are not allowed to dance together for a given number of dances right after a common dance.

In accord with these social rules, which can be quantified precisely based on empirical evidence[11], male and female dancers have to be understood and modelled as complex nodes with internal degrees of freedom. They are chosen as *body height, talent/vocation, activity* and *dancing skill*. Body heights are chosen based on statistical data for the german population[12], defining a suitably normalized gaussian distribution from which the dancers are randomly drawn. Note that the actual size distribution of the dancers thus is not a gaussian. Body height is a static feature of the dancer nodes, as are talent and activity, which all remain fixed after random attribution to the nodes. Depending on the talent for the dance, male and female dancers learn more or less rapidly. Activity models the intrinsic 'drive' of the dancer to participate in the dance and ask for partners. In accord with the different roles of male and female dancers, this variable is only set for males: we disregard females asking males for the dance. Finally, dancing skill S is a dynamic variable that changes (grows) in the course of the dancing events, i.e. $S = S(t)$. Skill evolves according to learning rules which differ for male and female dancers, again because of their different social roles. For female dancers, skill rank improves proportional to the number of dances with dancers of a given skill, while for male dancers, not the absolute skill of the female partner matters, but the respective skill difference. This allows males to improve their skill quite generally by practicing rather than by dancing only with optimal partners.

We have implemented these model features in an agent-based simulation of the tango argentino network, the code written in Java[11,13,14]. We here present simulation results on the tango network evolution for a system size of $N = 60$, with $F = M$. As adherence to the network is strong, we neglect fluctuations in the participant numbers; they are practically small and have no significant effect. The dance groups are chosen randomly from a representative gaussian distribution for the body heights, and vocation levels are assigned likewise. Finally, skill levels are set to an initial default value of 0. The simulation is then started, and the network of interactions starts to build up. We chose to look at about 50 dance events, each of which having about 16 dancing slots, which is a realistic size for a network within which no exchange of members occurs.

First, we characterize the network by the time evolution of the distribution of out-degrees, P(k), as shown in Figures 1 A and B for two different initial distributions of dancers. As the tango network model depends on several (empirical) parameters, a strictly universal behaviour can not be expected to hold for out system. In particular we note that expecting scale-free behaviour of the degree distribution would even be meaningless to expect (and also the sometimes observed algebraic behaviour of P(k) within one order of magnitude of values, sometimes observed and mistakenly called 'scale-free', does not occur in our system). For our system, we nevertheless find a generic behaviour in the shape of the degree distributions. Independently of the group of individuals chosen, the degree distributions quickly (i.e., at about 10 dance events) assume profiles which fluctuates only very little further on in the course of the dancing events. However, the shape of these final distributions remains strongly tied to the initial distribution, as the figures exemplify. Thus the whole network acquires an 'individualistic' character trait which can be predicted by simulations over a short time span only.

We have tested the robustness of these 'network individualistic' distributions with respect to variations of partner selection rules. Varying, e.g., the selection rule d) by increasing the waiting time of two partners between two dances leads to slightly more spread out degree distributions (not shown), as the dancers have to redistribute their dances to other network members. However, this does not affect the overall the emergence of a robust or stationary shape of the degree distribution. This also holds for variations according to rule c), i.e., the allowed number of common dances for one couple. The robustness of the network behaviour can further be corroborated by a computation of the clustering coefficient. After the initial transient it quickly enters a linear time regime in which it remains for the rest of the dancing event. This effect is most clearly demonstrated by normalizing the correlation coefficient to the mean out-degree (Figure 2). This quantity fully saturates already after the initial dancing period, despite significant individual variation of size and vocations (see insert), and a variation of two social rules (c and d). The saturation level, however, again strongly depends on the initial configurations.

We now turn to ask how far the social rules influence the individual's performance, i.e. how the connectivity and skill of individual changes within a given network. Individual activities can easily be visualized by using an array format in which male and female dancers are grouped in a 2-by-2 matrix (not shown). This leads to a time-series problem resembling to gene expression analysis of cDNA-microarrays, however, without any of the profound noise problems of the latter. A time series analysis of the distribution of skill levels of the individuals network members based on such arrayed representation formats can be summarized as follows. For male dancers, the skill evolution is dominated by their vocation or activity factors, rather than social rules. For both male and female dancers, we find a strong correlation between size and skill evolution, disfavouring both above and below average body heights. These individual skill evolution factors are discernable against the background of an overall linear growth in skill for the whole network group, which is a consequence of our skill evolution model.

The dynamics of individual skill evolution can most prominently be demonstrated by looking at the skill evolution of a new network member. We focus on a new female dancer (called the mentee in what follows), being introduced into the network by a girl friend, which is the most common situation. The girl friend serves as a mentor, and does so by transferring her existing connections to male dancers to the mentee. Assuming that always all existing connections are transferred we model the mentee as a connectivity-copy or -clone of the mentor. The mentoring process then falls into the class of processes wherein network growth is determined by strict node duplication (similar to the gene duplication process e.g. in protein networks[1]). Nodes of mentors and mentees, however, differ by their initial internal degrees of freedom, i.e. size, vocation and skill. In our simulation, we chose to select an existing network member (called the parent), and cloned it by maintaining its individual characteristics. Embodying it with the connections of the mentor, the resulting mentee thus is a hybrid of the individual characteristics of the parent and of the connectedness of the mentor.

For such a designed situation we have monitored the skill evolution of parent, mentor and mentee acting within a network of established connections (i.e. after the initial transient). Figure 3 A shows the result of such a simulation whereby different choices for the 'quality' of mentors and parents were made and held fixed. The quantities that were monitored are defined in Figure 3 B. We chose the distances of mentor, mentee and parent to the mean in skill distribution as 'order parameters', i.e. $d_x$ $S_{mean}(t) - S_x(t)$, where x is mentor (mr), mentee (me) or parent (pt). As can be seen in Figure 4 A, quite generally the success of the mentee, as measured by a decreasing value of $d_{me}$, depends significantly on the individual characteristics of the parent, but even more strongly on the connections acquired from its mentor. Consider, e.g., the case of a strong mentor, as shown in the left column. The evolution of the mentee is strongly coupled to the evolution of the mentor's skill, with the parent's influence mattering much less. For a poor parent and only an average mentor this still holds true, as seen in the bottom graph in the middle column. Thus, for our model social network we can say that the prospected skill evolution of the mentee most significantly depends on the quality of the mentor and less on those of the individual which, however, affect the speed of success.

To conclude, we have demonstrated our approach allows to study social networks with respect to both collective and individualistic behaviour. Individuality of members leads to individual system traits, but in a generic fashion. The effects of mentoring of newcomers in the model network point to the respective relevance of individual and mentor-related influences. It would be interesting to build on our approach to develop guided predictions for the success of individuals in their networks of interest, with appropriate modelling of the relevant features of both community and individual.

**Figure Captions**

Figure 1: Out-degree distributions P(k,t) and dance number distribution P(D) of the tango argentino network for the female dancers. The corresponding distributions for the males behave likewise. Note that the original histogram data have been systematically smoothed for better readability.

A) Out-degree distribution P(k) at three different times t, as measured in units of dance events. At early times, the distribution still fluctuates strongly, but at about t = 10 it settles into a rather stable, but still slightly fluctuating shape.

B) Same as in A), but for a different initial configuration of dancers.

C) Distribution of the total number of dances, P(D), for the dancers from A). Similar to P(k), P(D) assumes a rather stable profile. The number of dances is normalized to $D_{max}(t)$ to account for the increase in the total number of dances.

Figure 2: The time evolution of the clustering coefficient, normalized to the mean out-degree z = <k>, for simplicity ignoring the bipartiteness of the network graph. After the initial transient it assumes a stationary value, indicating that the network has reached a stationary state. The figure also shows that the effect of a variation of strategies on the network evolution is small: the colour-code refers to changes in the required waiting time between dances, and the allowed number of dances per partner. Different initial configurations display analogous behaviour, however, with different values of C/z (not shown). Insert: distribution of individual characteristics of the male and female dancers.

Figure 3: A) Skill evolution as measured by the order parameters $d_x$ for x = pt (red), mr (black), me (blue), by varying the 'quality' of the mentor in horizontal, and that of the parent in vertical direction. Parents were selected randomly from the skill distribution of the network members (after partitioning the distribution into three equal-sized parts). Mentors are the network members corresponding to the 25, 50 and 75 percentile.

B) Definition of order parameters $d_x$ for the mentor, mentee and parent. They are characterized by their respective distances to the mean (+) in distribution.

**Figure 1 A,B,C**

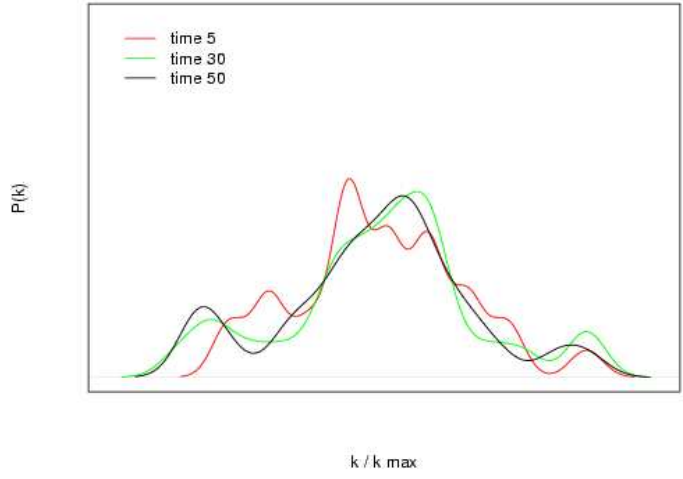

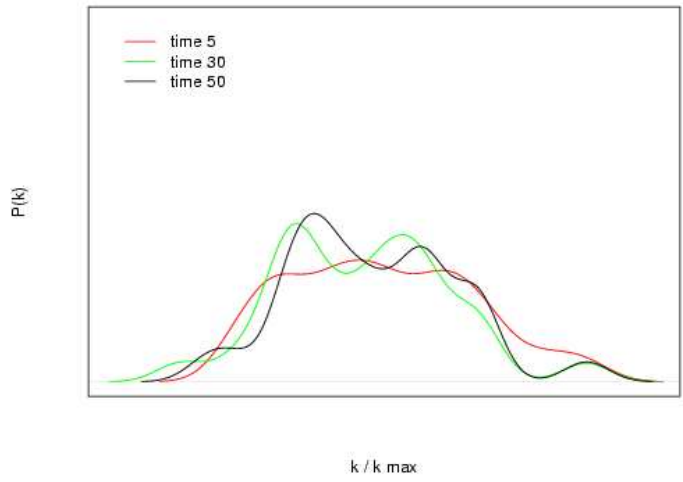

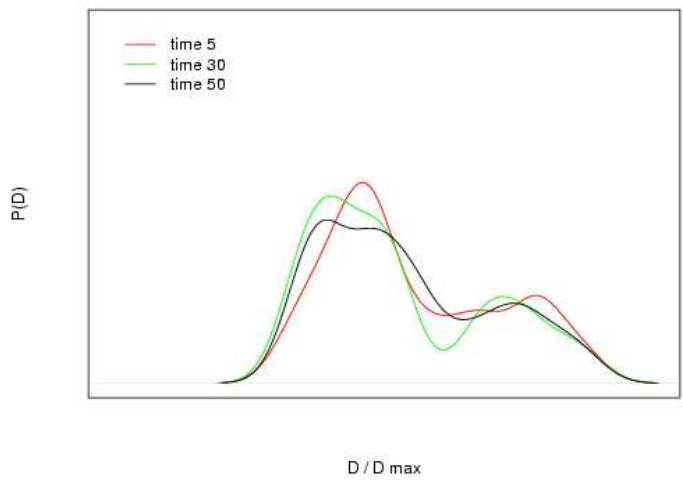

**Figure 2**

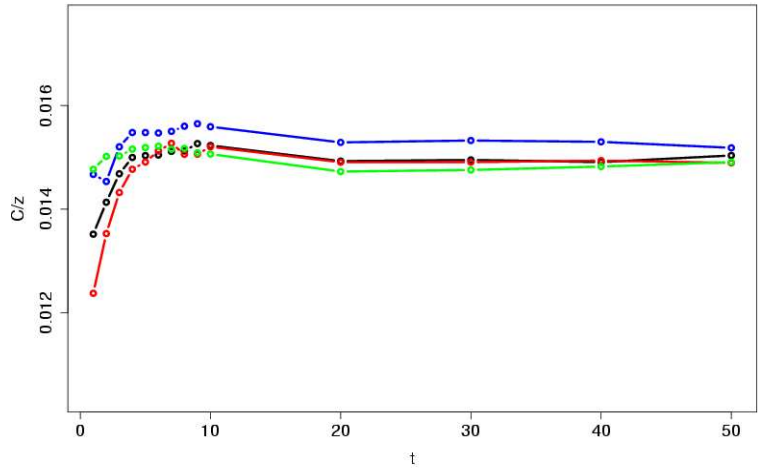

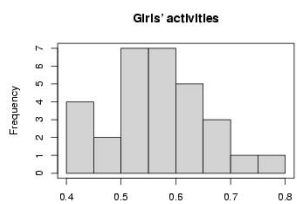
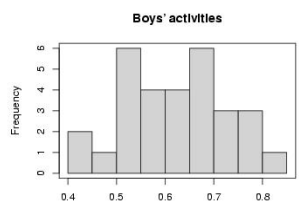

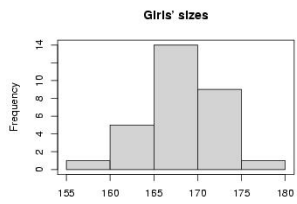
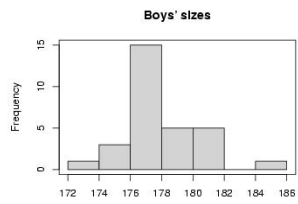

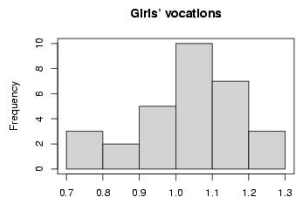
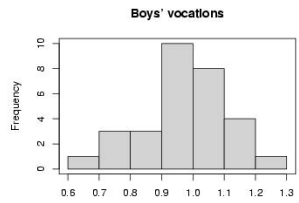

**Figure 3 A)**

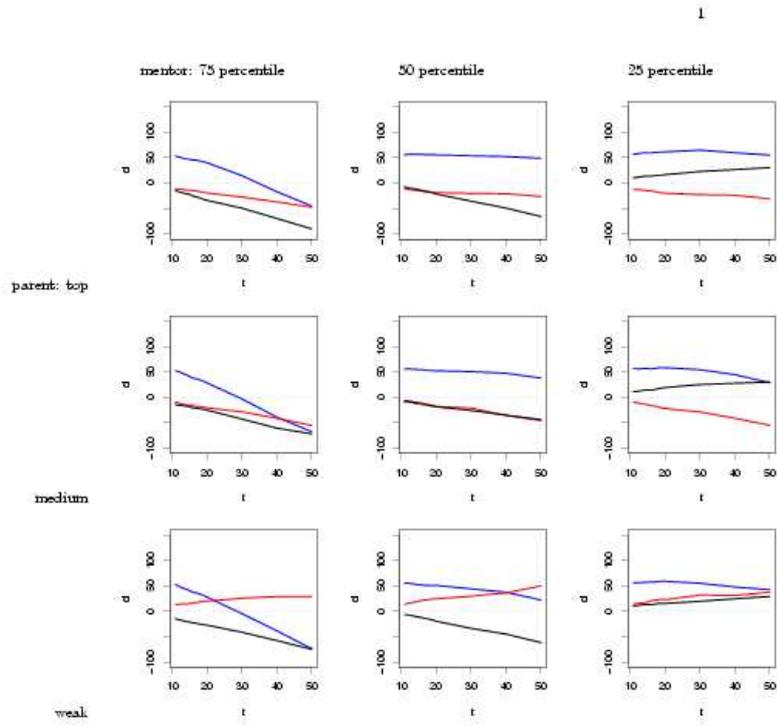

**Figure 3 B)**

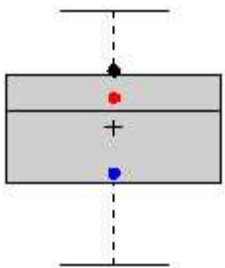